\documentclass[12pt]{article}
\usepackage{amsfonts,amssymb}
\newcommand{\dd}{ \hbox{d} }

\begin{document}

\title{\bf A Study of Brane Solutions in\\
$D$-dimensional Coupled Gravity System}
\author{Bihn Zhou\thanks{E-mail: zhoub@itp.ac.cn}\, and Chuan-Jie Zhu\thanks{
E-mail: zhucj@itp.ac.cn} \\
Institute of Theoretical Physics, Chinese Academy of\\
Sciences, P. O. Box 2735, Beijing 100080, P. R. China}

\maketitle

\begin{abstract}
In this paper, we use only the equation of motion for an interacting
system of gravity, dilaton and antisymmetric tensor to study the soliton
solutions by making use of a Poincar\'e invariant ansatz. We show that
the system of equations are completly integrable and the solution is
unique with appropriate boundary conditions. Some new class of solutions
are also given explicitly.
\end{abstract}

\section{Introduction}

Superstring theory \cite{GSW, Joe} was the leading candidate for a theory 
unifying  all matter and forces (including gravity, in particular). 
Unfortunately, there are five such consistent theories, namely, $SO(32)$
type I , type IIA, type IIB, $E_8 \otimes E_8$ heterotic and $Spin(32)/Z_2$
heterotic theories. This richness is an embarrassment for pure theorists.
One hope is that all these five superstring theories are just different
solutions of an underlying theory. In the last few years this hope turns
out to be true.  The underlying theory is the so called M theory
\cite{Hull, Wittena, Schwarz}. One  formulation of M theory is given in
terms of Matrix theory \cite{BFSS}. Nevertheless this formulation is not
background independent. It is difficult to use it to discuss non-perturbative
problems and still we should rely on the study of BPS states. A through
study of BPS states in any theory is a must for an understanding of
non-perturbative phenomena.  

In superstring theory and M theory, there is a plethora of BPS states.
These BPS states have the special property of preserving some supersymmetry.
In a low energy limit they are special solutions of the low energy
supergravity theory with Poincar\'e invariance. However there exists also
some other $p$-branes which don't preserve any supersymmetry and they are
also no longer Poincar\'e invariant. In finding these (soliton) solutions
one either resort to supersymmetry or make some simple plausible
assumptions. There is no definite reasoning that the so obtained solution
is unique. The purpose of this paper is to fill this gap.

It is also interesting to study soliton solutions for its own sake without
using any supersymmetric argument. Quite recently there are some interests
\cite{Klebanov, Minahan, Ferretti, Bergman} in studying branes in type 0
string theories which have no fermions and no supersymmetry in 10 dimensions
\cite{Joe}.  In these string theories, we can't use supersymmetric arguments.
So it is important to push other symmtric arguments to their limits.

In this paper we will study the coupled system of gravity ($g_{MN}$), dilaton
($\phi$) and anti-symmetric ($A_{M_1M_2\cdots M_{n-1}}$)
tensor in any dimensions. After making a Poincar\'e invariant ansatz for
the metric and either electric or magnetic ansatz for the  anti-symmetric
tensor, we derive all the equations of motion in some details.  A system
of five ordinary differential equations are obtained for four unknown
functions (of one variable). After making some changes of  the 
unknown functions we solved three equations explicity.  We show that the
remaining two equations (for one unknown functions) are mutually 
compatible and a unique solution can always obtained with appropriate
boundary conditions.  To solve the last equation in the most generic case
is surely beyond our ability because this last equation is a Riccati
equation which is known to be not solvable by algebric means. Instead
we will discuss some degenerate cases and recover many known solution
in the literature. Some new solutions are also found and their
physical property will be discussed elsewhere \cite{Zhou}. For previous
studies of soliton and brane solution in supergravity and string theories,
see for example the reviews \cite{Duff, Stelle, Argurio, Abers, Fre}.

The extension of the result of this paper to black branes will appear in 
another publication \cite{ZhouZhu}. We remark that this paper contains
some unnecessary computational details from some experts' point of view.
Although some formulas are not new (but scattered in various places),
we still think that they are necessary for an understanding of our logic. 

\section{The Equations of Motion and the Ansatz}

Our starting point is the following action for the coupled system of gravity
$g_{MN}$, dilaton $\phi$ and anti-symmetric tensor
$A_{M_1\cdots M_{n-1}}$:
\begin{equation}
I = \int \dd^D x \sqrt{- g} \, 
\left( R - { 1\over 2} g^{MN} \partial_M \phi \partial_N \phi -
{1 \over 2\cdot n! } e^{a\,\phi} F^2 \right),
\label{Action}
\end{equation}
where $a$ is a constant and $F$ is the field strength: $F=\dd A$.  An overall
proportional constant for the action $I$ is irrelevant for what follows. 

The equations of motion can be easily derived from the above action
(\ref{Action}):
\begin{eqnarray}
  & & R_{MN} = \frac{1}{2} \partial_M\phi\partial_N\phi + S_{MN},
\label{Req} \\
  & & {1\over \sqrt{-g} }\partial_{M_1} ( \sqrt{-g} \,
e^{a\phi} F^{M_1 M_2 \cdots M_n})  = 0,
\label{Feq} \\
  & & {1\over \sqrt{-g} } \partial_M( \sqrt{-g} g^{MN} \partial_N\phi)  
= {a\over 2 n!}\, e^{a\phi}\, F^2,
\label{phieq}
\end{eqnarray}
where
\begin{equation}
  S_{MN} = {1\over 2(n-1)!} e^{a\phi}\, \left(F_{M M_2 \cdots M_n}
	F_N^{\ M_2 \cdots M_n} -{n-1\over n(D-2)} F^2 g_{MN}\right).
\end{equation}
Of course it is impossible to solve the above system equations in their full
generality. To get some meaningful solution we will make some assumptions
by using symmetric arguments.

Our ansatz for a $p$-dimensional brane is as follows:
\begin{equation}
\dd s^2 = e^{2 A(r)}(- \dd t^2 + 
\sum_{\alpha=1}^p (\dd x^{\alpha})^2 ) +
e^{ 2 B(r)} \dd r^2 +
e^{ 2 C(r)} \dd \Omega_{\tilde{d}+1}, \label{ansatz}
\end{equation}
where $\dd \Omega_{\tilde{d}+1}$ is the square of the line element on the unit
$\tilde{d}+1$ sphere which can be written as follows ($D=p+\tilde{d}+3$):
\begin{equation}
\dd \Omega_{\tilde{d}+1} = \dd \theta_1^2 + \sin^2 \theta_1 \dd \theta_2^2
+ \cdots + (\sin\theta_1 \cdots \sin\theta_{\tilde{d}})^2
\dd\theta^2_{\tilde{d}+1}.
\end{equation}
The brane is extended in the directions $(t, x^{\alpha})$. The ansatz
(\ref{ansatz}) is Poincar\'e invariant in these directions. For our later
use in \cite{ZhouZhu} we will give the formulas for the more general
ansatz:
\begin{equation}
\dd s^2 = - e^{2 A(r)} \dd t^2 + 
\sum_{\alpha=1}^p e^{2 A_{\alpha}(r)} (\dd x^{\alpha})^2+
e^{ 2 B(r)} \dd r^2 +
e^{ 2 C(r)} \dd \Omega_{\tilde{d}+1}.
\end{equation}
We also note here that there is still some freedom to choose the
parametrization $r$. This freedom can be fixed by making either the choice
$C(r)=\ln r$ or $C(r) =B(r) + \ln r$.

For the anti-symmetric tensor $A$, we have 2 different choices. The first
choice is the electric case where we take the following form for $A$
($x^0 =t$):
\begin{equation}
A =\pm  e^{\Lambda(r)}\, \dd x^0\wedge \dd x^1 \wedge \cdots \wedge \dd x^p.
\end{equation}
This fixes $p = n-2$. 
The second choice is the magnetic case where we take the following form for
the dual potential $\tilde{A}$:
\begin{equation}
\tilde{A} = \pm
e^{\Lambda(r)}\, \dd x^0\wedge \dd x^1 \wedge \cdots \wedge \dd x^p,
\label{ansatzb}
\end{equation}
and $p = D-n-2$.
We note that the relation between the antisymmetric tensor field strength $F$
and its dual field strength $\tilde{F}$ ($\equiv \dd \tilde{A}$) is:
\begin{equation}
F^{M_1\cdots M_n} = { 1\over \sqrt{ -g } } \, e^{ -a \phi} \, 
{ 1\over (D-n)! } \,
\epsilon^{M_1 \cdots M_n N_1 \cdots N_{D-n} } \, \tilde{F}_{
N_1 \cdots N_{D-n} } . 
\end{equation}
By using this relation the ansatz (\ref{ansatzb}) transform to an ansatz
for $F$:
\begin{equation}
F = \pm 
\Lambda' \hbox{exp}( \Lambda - a \phi -A-\sum_{\alpha=1}^p A_{\alpha}
-B + (\tilde{d}+1)C) \, \omega_{\tilde{d}+1}, 
\end{equation}
where $\omega_{\tilde{d}+1}$ is the volume form of the sphere
$S^{\tilde{d}+1}$ with unit radius.
As it is well-known from duality, the equation of motion (\ref{Feq}) becomes
the Bianchi identity for $\tilde{F}$ which is satisfied automatically.
Nevertheless the Bianchih identity for $F$ becomes the equation of motion
for $\tilde{F}$  which is given as follows:
\begin{equation}
{1\over \sqrt{-g} }\partial_{N_1} ( \sqrt{-g} \,
e^{-a\phi} \tilde{F}^{N_1 N_2 \cdots N_{D-n}})  = 0.
\label{Feqa}
\end{equation}
From now on we will discuss the electric case only.

We will compute the Riemann curvature by using the vilebein formalism.
The coordinate index is denoted as $(M) = (t, \alpha, r, i)$ and tangent
index is denoted as $(A)=(\bar{t},\bar{\alpha},\bar{r}, \bar{i})$, 
i.e. a bar over coordinate index. The moving frame is:
\begin{eqnarray}
e^{\bar{t}} & = & e^{A(r)} \, \dd t = - e_{\bar{t}} , 
 \\
e^{\bar{\alpha}} & = & e^{A_{\alpha}(r) } \,  \dd x^{\alpha} = e_{\bar{\alpha}},
\\
e^{\bar{r}} & = & e^{ B(r)} \, \dd r = e_{\bar{r}}, 
\\
e^{\bar{i}} & = & e^{ C(r)} \, \sin\theta_1 \cdots \sin\theta_{i-1}\dd
\theta_{i}  = e_{\bar{i}} .
\end{eqnarray}
By using the defining equation of $\omega^{AB}$:
\begin{equation}
\dd e^A + \omega^{AB} \wedge  e_B = 0, 
\end{equation}
we obtain the following expression for $\omega^{AB}$:
\begin{eqnarray}
\omega^{ \bar{t}\bar{r}} & = & A'\, e^{A-B}\dd t = A'\, e^{-B} \, e^{\bar{t}},
\\
\omega^{ \bar{\alpha}\bar{r}} & = & A'_{\alpha} \, e^{A_{\alpha} -B} \dd
y_{\alpha} = A'_{\alpha} \, e^{-B }\, e^{\alpha}, \\
\omega^{\bar{i}\bar{r}} & = & C'\,e^{C-B} \,\sin\theta_1 \cdots
\sin\theta_{i-1}\dd \theta_{i}  = C'\,e^{-B} \, e^{\bar{i}}, \\
\omega^{\bar{i}\bar{j}} & = & \cos\theta_j \sin\theta_{j+1} \cdots \sin
\theta_{i-1} \dd \theta_i , \qquad i>j. 
\end{eqnarray}
The other components are zero or can be obtained from the above by using
the anti-symmetric property of $\omega^{AB}$:~~$\omega^{AB}= - \omega^{BA}$.

From $\omega^{AB}$ we can compute $R^{AB}$:
\begin{equation}
R^{AB} = \dd \omega^{AB} + \omega^{AC}\wedge \eta_{CD}\omega^{DB}.
\end{equation}
The results are:
\begin{eqnarray}
R^{\bar{t}\bar{\alpha} } & = & - A'A'_{\alpha} \, e^{-2 B} \,e^{\bar{t}}
\wedge e^{\bar{\alpha}} , \\
R^{\bar{t}\bar{r} } & = & -(A'e^{A-B})'\, e^{-A-B} \, e^{\bar{t}}\wedge
e^{\bar{r}}, \\
R^{\bar{t}\bar{i}} & = & - A'C' \, e^{-2B} \, e^{\bar{t}}\wedge e^{\bar{i}}, \\
R^{\bar{\alpha}\bar{\beta}} & = & - A_{\alpha}'A_{\beta}' \, e^{-2 B} \, 
 e^{\bar{\alpha}}\wedge e^{\bar{\beta}}, \\
R^{\bar{\alpha}\bar{r} } & = & -(A_{\alpha}'e^{A_{\alpha} -B})' \, 
e^{-A_{\alpha}-B} 
 e^{\bar{\alpha}}\wedge e^{\bar{r}} , \\
R^{\bar{\alpha}\bar{i}} & = & - A_{\alpha}'C' \, e^{-2B} \, e^{\bar{\alpha}}
\wedge e^{\bar{i}}, \\
R^{\bar{r}\bar{i}} & = & -(C'e^{C-B})' \, e^{-C-B} \, e^{\bar{r}}\wedge
e^{\bar{i}}, \\
R^{\bar{i}\bar{j}} & = & (e^{-2C} - (C')^2\, e^{-2B}) \, 
 e^{\bar{i}}\wedge e^{\bar{j}}. 
\end{eqnarray}
We note that all $R^{AB}$'s are of the following special form:
\begin{equation}
R^{AB} = f^{AB} \, e^{A}\wedge e^{B}, 
\end{equation}
where there is no summation over $A$ and $B$.
$f^{AB}$ can be chosen to be symmetric and is 0
when $A=B$. By definition 
\begin{equation}
R^{AB} = {1\over 2} \, R^{AB}_{MN} \, \dd x^M\wedge \dd x^N, 
\end{equation}
we have 
\begin{equation}
R^{AB}_{MN} = f^{AB}(e^A_M\, e^B_N - e^A_N \, e^B_M), 
\end{equation}
and 
\begin{equation}
R^A_M= R^{AB}_{MN}\, e^N_B = \sum_B f^{AB} \, e_M^A \equiv f^A \, e^A_M,
\end{equation}
where we have defined $f^A$ as
\begin{equation}
f^A=\sum_B f^{AB}, 
\end{equation}
which can be computed to give the following results:
\begin{eqnarray}
f^{\bar{t}} & = & -e^{-2B}\Big( A'' + A'(A' + \sum_{\alpha} A_{\alpha}' - B' +
 ( \tilde{d} + 1 )C')\Big), 
\\
f^{\bar{\alpha}} & = & - e^{-2 B} 
\Big( A_{\alpha}'' + A_{\alpha}'(A' + \sum_{\beta} A_{\beta}' - B' +
 (\tilde{d} + 1 )C')\Big), 
\\
f^{\bar{r}} & =& -e^{-2B} \Big( A'' + \sum_{\alpha}A_{\alpha}'' + 
(\tilde{d} + 1 )C'' +(A')^2 + 
\sum_{\alpha}(A_{\alpha}')^2 \nonumber \\
&& + (\tilde{d} + 1 )(C')^2 -B'(A' + \sum_{\alpha} A_{\alpha}' +
 (\tilde{d} + 1 )C')\Big) , 
\\
f^{\bar{i}}  & = & - e^{-2B}\, \Big( C'' + C'(A' + \sum_{\alpha} A_{\alpha}'
-B'+ (\tilde{d} + 1 )C')
\nonumber \\
& & - \tilde{d} \, e^{-2C +2B}\Big).
\end{eqnarray}
By making use of fact that the metric in (\ref{ansatz}) is a diagonal metric
we have \begin{eqnarray}
R_{MN} & = & R^A_M \, e_{AN} = f^A\delta^A_M \, g_{MN},
\\
R_{MN}\dd x^M \otimes \dd x^N & = & 
- f^{\bar{t}} \, e^{\bar{t}} \otimes e^{\bar{t}} + 
\sum_{\alpha=1}^p f^{\bar{\alpha}} \, e^{\bar{\alpha}} \otimes e^{\bar{\alpha}}
\nonumber \\
& & +  f^{\bar{r}} \, e^{\bar{r}} \otimes e^{\bar{r}} + f^{\bar{i}}
\dd \Omega_{\tilde{d}+1},
\end{eqnarray}

To obtain the equations of motion from (\ref{Req})--(\ref{phieq}) we
also need to know some expressions 
involving $F$. We have (for electric solution and $p=n-2$)
\begin{eqnarray}
A & = & \pm e^{\Lambda(r)} \, \dd x^0 \wedge \dd x^1 \cdots \wedge \dd x^{n-2}, 
\\
F & = & \dd A = \pm \Lambda'\,e^{\Lambda(r)} \, \dd r \wedge \dd x^0 \wedge 
\dd x^1 \cdots \wedge \dd x^{n-2}, 
\\
F_{MN} & \equiv  & F_{MM_2\cdots M_n} {F_N}^{M_2\cdots M_n} , \\
F_{rr} & = & -(n-1)! \, (\Lambda'\,e^{\Lambda(r)} )^2 \, e^{ -2(n-1)A} , \\
F_{\mu\nu} & = & -\eta_{\mu\nu}\,(n-1)! \, (\Lambda'\,e^{\Lambda(r)} )^2 \, 
e^{ -2(n-2)A-2B}, \\
F_{MN} & = & 0, \qquad \hbox{for the rest cases,}
\end{eqnarray}
and 
\begin{equation}
F^2 = g^{MN}\, F_{MN} = - n!\, (\Lambda'\,e^{\Lambda(r)} )^2 \, 
e^{ -2(n-1)A-2B}, 
\end{equation}

The equation of motion for $g^{MN}$ gives the following equations:
\begin{eqnarray}
 f^{\bar{r}} g_{rr} & = &{1\over 2}(\partial_r \phi)^2 + { 1\over 2\cdot
(n-1)!}\, e^{a \,\phi} \left(F_{rr} - {(n-1)\over n(D-2) }\, F^2 \, g_{rr}
\right), 
\\
 f^{\bar{\mu}}\, g_{\mu\nu} & = &
{ 1\over 2\cdot (n-1)!}\, e^{a \,\phi} \left(F_{\mu\nu} -
{(n-1)\over n(D-2) }\, F^2 \, g_{\mu\nu} \right), 
\\
 f^{\bar{i}} g_{ij} & =& { 1\over 2\cdot (n-1)!}\,
 e^{a \,\phi} \left(F_{ij} - {(n-1)\over n(D-2) }\, F^2 \, g_{ij} \right),
\end{eqnarray}
for the $(rr)$, $(\mu\nu)$ and $(ij)$ componenets respectively. Substituting all
$f^{A}$ and $F_{MN}$ into the above equations and setting $C =B + \ln r$,
$A_{\alpha} =A$, 
we obtain the following three equations:
\begin{eqnarray}
& & A'' + d  (A')^2 + \tilde{d}\, A'B' + { \tilde{d}+1\over r } \, A' =  
{ \tilde{d}\over 2 (D-2) } \, S^2 , 
\label{lasta}\\
& & B'' + d  A'B' + {d \over r} \,A' +\tilde{d} (B')^2 
\nonumber \\
& & \hskip 3cm + { 2 \tilde{d} + 1 \over r}\, B' = - { 1\over 2}\, 
{n-1\over D-2}\, S^2, 
\\
& & d A'' + (\tilde{d} + 1)B'' + d (A')^2 + { \tilde{d} + 1 \over r}\, B'
\nonumber \\
& & \hskip 3cm  - d A'B' + { 1\over 2} \, (\phi')^2 = 
{1\over 2}\, {\tilde{d} \over D-2}\, S^2,
\label{lastb}
\end{eqnarray}
where $S = \Lambda' \, e^{ {1\over 2}\,  a\phi + \Lambda - d \, A }$ and
$d = p+1 = n-1$. The equation of motion for $\phi$ is
\begin{equation}
\phi'' + \Big( d A' + \tilde{d}B' + {\tilde{d} + 1 \over r} \Big) \phi
= - { a\over 2} \, S^2 ,
\label{lastc}
\end{equation}
and the equation derived from the equation of motion for $F$ is
\begin{equation}
(\Lambda'(r) \, e^{\Lambda(r) + a \phi(r) - d  A(r) + 
\tilde{d}  B(r) } \, r^{\tilde{d}+1} ) '= 0.
\label{lastd}
\end{equation}
These five equations, eqs. (\ref{lasta})-(\ref{lastb}), (\ref{lastc}) and
(\ref{lastd}), consist of the complete system of equations of motion for
four unknow functions: $A(r)$, $B(r)$, $\phi(r)$ and $\Lambda(r)$.  We will
discuss these equations in the next two sections.

\section{The Solvability of the Equations}

In this section we will try to solve the system of the above five equations.
First it is easy to integrate eq. (\ref{lastd}) to get 
\begin{equation}
\Lambda'(r) \, e^{\Lambda(r) + a \phi(r) - d  A(r) + 
\tilde{d}  B(r) } \, r^{\tilde{d}+1}  = C_0,
\label{czero}
\end{equation}
where $C_0$ is a constant of integration. If we know the other three functions
$A(r)$, $B(r)$ and $\phi(r)$, this equation can be easily integrated to give 
$\Lambda(r)$:
\begin{equation}
e^{\Lambda(r)} = C_0\, \int^r \dd r \, { e^{ - a \phi(r) +  d  A(r) - \tilde{d}
B(r)  } \over  r^{\tilde{d} + 1 } } . 
\end{equation}
By using eq. (\ref{czero}), $S$ can be written as 
\begin{equation}
S(r) = C_0 { e^{ - {a \over 2 } \phi(r) - \tilde{d}  B(r) } \over r^{\tilde{d}
 +1} } . 
\label{Sczero}
\end{equation}

Now we make a change of functions from $A(r)$, $B(r)$ and $\phi(r)$ to 
$\xi(r)$, $\eta(r)$ and $Y(r)$:
\begin{eqnarray}
\xi(r) & = &  d A(r) + \tilde{d} B(r),
\\
\eta(r) & = & \phi(r) +   a\big( A(r)-B(r) \big),
\\
Y(r) & = & A(r)-B(r),
\end{eqnarray}
or 
\begin{eqnarray}
A  & = & { \xi + \tilde{d} Y \over d + \tilde{d} }, 
\\
B  & = & { \xi -  {d} Y \over d + \tilde{d} }, 
\\
\phi  & = & \eta - a \, Y.  
\end{eqnarray}
The equations  are then changed to
\begin{eqnarray}
& &\xi^{\prime\prime}+(\xi^\prime)^2+\frac{2\tilde{d}+1}{r} \xi^\prime
  =0,
\label{xieq}  \\
& &\eta^{\prime\prime}+\left(\xi^\prime+\frac{\tilde{d}+1}{r}\right)\eta^\prime
  -\frac{ a}{r}\xi^\prime =0,
\label{etaeq}  \\
& &Y^{\prime\prime}-\frac{\Delta}{2}(Y^\prime)^2+\left(\frac{\tilde{d}-d}{D-2}
  \xi^\prime+\frac{\tilde{d}+1}{r}+ a\, \eta^\prime\right)Y^\prime
\nonumber \\
&  & \hskip 2cm  -\frac{1}{2}(\eta^\prime)^2-\xi^{\prime\prime}+\frac{1}{D-2}
  (\xi^\prime)^2 =0,
\label{Yeq}  \\
&  & Y^{\prime\prime}+\left(\xi^\prime+\frac{\tilde{d}+1}{r}
  \right) Y^\prime-\frac{1}{r}\xi^\prime = \frac{1}{2}S^2,
\label{newSeq}
\end{eqnarray}
where
\begin{equation}
\Delta = \frac{2d\tilde{d}}{D-2} + a^2.
\end{equation}

The general solutions  for $\xi$ and $\eta$  can be obtained easily from
eqs. (\ref{xieq}) and (\ref{etaeq}) and we have
\begin{eqnarray}
\xi  & = &\ln \left| C_1 + C_2 r^{-2\tilde{d}} \right|,
\label{sumdat} \\
\eta^\prime  & = & \frac{ 2C_2   a + C_3 r^{\tilde{d}} }
  {r(C_2 + C_1 r^{2\tilde{d}})},
\label{phfdat}
\end{eqnarray}
where $C_1$, $C_2$ and $C_3$ are constants of integration. To make
sense of the above expressions, $C_1$ and $C_2$ can't be zero
simultaneously. For $C_1\neq0$ we can always choose $C_1=1$ by
a rescaling of $t$ and $x^{\alpha}$.

Substituting the above expressions into eqs. (\ref{Yeq}) and (\ref{newSeq}),
we obtain
\begin{equation}
Y^{\prime\prime} - \frac{\Delta}{2}\, (Y^\prime)^2 + Q(r)Y^\prime = R(r)
\label{NDf}
\end{equation}
and
\begin{equation}
  S^2=\Delta\left( Y^\prime
  -\frac{ 2C_2+C_3\frac{  a}{\Delta}r^{\tilde{d}} }
  { r(C_2+C_1r^{2\tilde{d}})} \right)^2
  +\frac{K}{\Delta(C_2+C_1r^{2\tilde{d}})^2 }r^{2\tilde{d}-2},
\label{Srslt}
\end{equation}
where
\begin{eqnarray}
    Q(r) & = & \frac{\tilde{d}+1}{r}
    + \frac{ 2C_2(\Delta-\tilde{d}) + C_3   a r^{\tilde{d}} }
    {r(C_2 + C_1 r^{2\tilde{d}}) },
\label{Qdef}\\
R(r)&=&\frac{2C_2^2(\Delta-\tilde{d})+2C_2C_3  ar^{\tilde{d}} +2C_1C_2
\tilde{d}(2\tilde{d}+1) r^{2\tilde{d}} +\frac{1}{2}C_3^2 r^{2\tilde{d}} }
{ r^2(C_2 +C_1 r^{2\tilde{d}})^2 },
\label{Rdef}
\end{eqnarray}
where 
\begin{equation}
K=C_3^2(\Delta-a^2)+8C_1C_2\Delta\tilde{d}(\tilde{d}+1),
\end{equation}
is a constant. 

From eq. (\ref{Srslt})  we see that if $K \geq 0$, there is no
restriction for the interval of $r$. If the constant $K<0$, $r$ can only take
such values that where $S^2\geqslant 0$.

Notice that our system of equations of motion is an over determined system:
four unknown functions satisfying five equations. We have solved three 
equations and there are two equations, eqs. (\ref{NDf}) and (\ref{Srslt}), 
remaining with one unknown functions $Y(r)$. Now we would like to show that
these two equations actually give no constraints on $Y(r)$ and effectively
there is only one equation, i.e., we can solve either one of them and the
other one will be satisfied automatically.

Because this proof is a rather involved algebraic calculation which we did
it by using Mathematica using computer, we will only give the steps of the
proof.

First we change eq. (\ref{Srslt}) by taking the logarithm of both sides. The
logarithm of $S$ can be rewritten as a function of $Y$, $\xi$ and $\eta$ by
using eq. (\ref{Sczero}). Now  we differentiate both sides with respect to
$r$ to eliminate $C_0$. Substituting $\xi$ and $\eta'$ with the explicit
results given in eqs. (\ref{sumdat}) and (\ref{phfdat}) we obtain an equation
containing  only the unknow function $Y$ and its first and second derivatives
$Y'$ and $Y''$.

In the second step we eliminate $Y''$ by using eq. (\ref{NDf}). We found that
the remaining equation involving $Y$ and $Y'$ is just an identity\footnote{
Otherwise we would obtain an equation containing $Y$ and $Y'$ only. It could
be used to solve $Y$.}. This completes our proof that eq. (\ref{NDf}) and
eq. (\ref{Srslt}) are compatible and we can use either one to solve for $Y$.
It seems that eq. (\ref{NDf})  is more suitable for solving $Y$.

Now we start to solve equation (\ref{NDf}). By a change of function from $Y$
to $y$:
\begin{equation}
y = Y' - { 1\over \Delta }\, Q(r),
\end{equation}
eq. (\ref{NDf}) becomes
\begin{equation}
y' - { \Delta \over 2 } \, y^2 = \tilde{R}(r), 
\label{lasty}
\end{equation}
which is known as a kind of Ricatti equation. Here
\begin{eqnarray}
  \tilde{R}(r) & = & R(r) -\frac{1}{\Delta} Q^\prime(r) - \frac{1}
	{2\Delta}Q^2(r)
  \nonumber \\
    & = & -\frac{\tilde{d}^2-1}{2\Delta r^2} + \frac{\tilde{\Lambda}}{2\Delta}
    \frac{r^{2\tilde{d}-2}}{ (C_2+C_1 r^{2\tilde{d}})^2 }
\label{NRdef}
\end{eqnarray}
with
\begin{eqnarray}
  \tilde{\Lambda} &=& K - 4C_1C_2\tilde{d}^2
\nonumber \\
  &=& C_3^2 (\Delta-a^2) + 4C_1 C_2
  \tilde{d}(2\tilde{d}\Delta + 2\Delta - \tilde{d}).
\end{eqnarray}

It is well-known that the above Ricatti equation can't be solved algebraically
for generic $\tilde{R}(r)$. If we know any special solution to (\ref{lasty}),
a general construction could be used to find the  geneal solution to eq.
(\ref{NDf}). Denoting the special solution to (\ref{lasty}) by $y_0(r)$, the
general solution to eq. (\ref{NDf}) is given as follows:
\begin{equation}
 Y(r) =\int^r y_0(r')\dd r' - \frac{2}{\Delta}\ln \bigg| C_4 + C_5 \int^r
  e^{\Delta\int^{r'} y_0(r'')\dd r''} \dd r'\bigg|
  +\frac{1}{\Delta}\int Q(r') \dd r',
\label{constrY}
\end{equation}
where $C_4$ and $C_5$ are two constants of integration.

In the next section we will solve some degenerate cases. 
In this way we got the well-known $p$-brane solution. 

\section{Some Degenerate Cases}

\subsection{Case of $\tilde{\Lambda}=0$}

In this case $\tilde{R}$ was reduced to
\begin{equation}
  \tilde{R}(r)=-\frac{\tilde{d}^2-1}{2\Delta r^2}.
\label{Rtilde}
\end{equation}
It is easy to find a special solution to eq. (\ref{lasty}):
\begin{equation}
  y_0 = \frac{\tilde{d}-1}{\Delta r}.
\end{equation}
By using eq. (\ref{constrY}) we have
\begin{eqnarray}
  Y& =& -\frac{2}{\Delta}\ln\left| C_4+C_5 r^{-\tilde{d}} \right|
  - \frac{\Delta-\tilde{d}}{\Delta\tilde{d}}
    \ln\left| C_1+C_2 r^{-2\tilde{d}} \right| 
\nonumber \\
& &  
  + \frac{C_3\, a}{2 \tilde{d} \Delta  \sqrt{-C_1C_2}  }
    \ln \left| {  \sqrt{-C_1C_2}   + C_2 \, r^{- \tilde{d}} \over 
 \sqrt{-C_1C_2}   - C_2 \, r^{- \tilde{d}} }\right| , 
\end{eqnarray}
where we have redefined $C_4$ and $C_5$.  From this we get
\begin{eqnarray}
  A(r)  &=& \frac{\tilde{d}}{\Delta(D-2)}
  \ln\left| C_1+C_2 r^{-2\tilde{d}}\over (C_4+C_5 r^{-\tilde{d}})^2 
\right| 
\nonumber \\
& & 
  + \frac{a \,C_3 }{2 \Delta(D-2)\sqrt{-C_1C_2}}
\ln \left| \sqrt{-C_1C_2} + C_2 r^{-\tilde{d}} 
\over \sqrt{-C_1C_2} - C_2 r^{-\tilde{d}}\right| ,
\\
  B(r) &=& \frac{\Delta+a^2}{2\Delta\tilde{d}}
  \ln\left|C_1+C_2 r^{-2\tilde{d}} \right|
  + \frac{2d}{\Delta(D-2)} \ln\left|C_4+C_5 r^{-\tilde{d}}\right|
  \nonumber \\
& & - \frac{C_3 \,a\, d}{2 \tilde{d} \Delta(D-2) \sqrt{-C_1C_2}    }
\ln \left| \sqrt{-C_1C_2} + C_2 r^{-\tilde{d}} 
\over \sqrt{-C_1C_2} - C_2 r^{-\tilde{d}}\right| ,
 \\
\phi(r)  &=& 
\frac{C_3\,d }{ \Delta (D-2) \sqrt{-C_1C_2} }
\ln \left| \sqrt{-C_1C_2} + C_2 r^{-\tilde{d}} 
\over \sqrt{-C_1C_2} - C_2 r^{-\tilde{d}}\right| 
\nonumber \\ 
& & 
  - \frac{ a}{\Delta}\ln\left|C_1+C_2 r^{-2\tilde{d}}\right|
  + \frac{2 a }{\Delta}\ln\left|C_4+C_5 r^{-\tilde{d}}\right|
  + \ln C_6,
\nonumber \\
  \Lambda'(r) \, e^{\Lambda} &=& { C_0 \, C_6^{-a} \over  
 r^{\tilde{d}+1}(C_4+C_5 r^{-\tilde{d}})^2}.
\end{eqnarray}
One can check explicitly that the other equation (\ref{Srslt})  is also
satisfied if we have
\begin{equation}
C_0= \pm 2\,\tilde{d}\,\sqrt{\frac{\left| C_1C_5^2+C_2C_4^2 \right|} 
  {\Delta} } \, \, C_6^{\frac{1}{2}\, a}.
\end{equation}

From the structure of the above solution it is consistent to set $C_3 = 0$.
Because of $\tilde{\Lambda} = 0 $, this requires $C_1C_2=0$. The two cases
$C_1=0$ or $C_2=0$ are acutally equivalent. They are related by a
$r\rightarrow 1/r$ symmetry which is discussed in \cite{Zhou}. So we set
$C_2=0$,  and $C_1=C_4=1$ by a conventional choice of scale. The solution
simplifies to:
\begin{eqnarray}
A(r)  & = & - { 2 \tilde{d} \over (D-2) \Delta} \, 
\ln \left(1 + C_5 \, r^{-\tilde{d}} \right), 
\\
B(r) & = &  { 2 {d} \over (D-2) \Delta} \, 
\ln \left(1 + C_5 \, r^{-\tilde{d}} \right), 
\\
\phi(r) & = &  { 2 a  \over \Delta} \, 
\ln \left(1 + C_5 \, r^{-\tilde{d}} \right), 
\\
e^{\Lambda(r)} & = & { 2\over \sqrt{\Delta} } 
\,{1 \over  \left(1 + C_5 \, r^{-\tilde{d}} \right)}.
\end{eqnarray}
This is the well-known BPS $p$-brane solution.

We will not discuss the physical meaning of these solutions here. For details
see \cite{Zhou} or  the reviews \cite{Duff, Stelle, Argurio, Abers, Fre}.

\subsection{Special Cases of $\tilde{\Lambda}\neq 0$}

In this subsection we will solve some special cases which didn't fall into the
case $\tilde{\Lambda}=0$.

\subsubsection{ $C_1=0$}

In this case  eq. (\ref{lasty}) becomes
\begin{equation}
  y^\prime-\frac{\Delta}{2}y^2=-\frac{\tilde{d}^2-1}{2\Delta r^2}
  +\frac{C_3^2(\Delta-a^2)}{2C_2^2\Delta} r^{2\tilde{d}-2}
\end{equation}
which has the following two special solutions:
\begin{equation}
  y_\pm=\frac{\tilde{d}-1}{\Delta r}\pm i\frac{C_3}{C_2}\
  \frac{\sqrt{\Delta-a^2}}{\Delta}r^{\tilde{d}-1}.
\end{equation}
Eventhough the above  solutions are complex, we can still obtain a real
solution. For details see \cite{Zhou}.

\subsubsection{$C_2=0$}

This case is  quite similar to the case $C_1=0$ discussed in the last
subsection. In fact there is a symmetry $r\rightarrow 1/r$  as mentioned
above \cite{Zhou}.

\subsubsection{}

So far we have discussed  all the possible solutions when either  $C_1$
or  $C_2$ is zero. When none of them is zero, things  become more
complicated. The interested reader can find more special cases in \cite{Zhou}
not discussed here.

\section *{Acknowledgments}
We would like to thank Han-Ying Guo, Yi-hong Gao, Ke Wu, Ming Yu, Zhu-jun
Zheng and Zhong-Yuan Zhu for discussions. This work is supported in part
by funds from Chinese National Science Fundation and Pandeng Project.

\end{document}